\documentclass[a4paper,11pt]{article}
\usepackage{jcappub}
\usepackage{float}
\usepackage{bm}
\usepackage[utf8]{inputenc}
\usepackage{amsmath}
\usepackage{amssymb}
\usepackage{subeqnarray}
\usepackage{graphicx}
\usepackage{xcolor}
\usepackage{siunitx}
\usepackage{subcaption}
\usepackage[normalem]{ulem}
\usepackage{ragged2e}
\usepackage{aas_macros}
\usepackage{geometry}
\usepackage{epsfig}
\usepackage{placeins}

\newcommand\spart{\;\raise1.0pt\hbox{/}\hskip-6pt\partial}
\newcommand\spartb{\;\overline{\raise1.0pt\hbox{/}\hskip-6pt\partial}}

\newcommand{\be}{\begin{equation}}
	\newcommand{\ee}{\end{equation}}
\newcommand{\bea}{\begin{eqnarray}}
	\newcommand{\eea}{\end{eqnarray}}
\newcommand{\beal}{\begin{align}}
	\newcommand{\eeal}{\end{align}}
\newcommand{\beas}{\begin{subeqnarray}}
	\newcommand{\eeas}{\end{subeqnarray}}
\newcommand{\dd}{{\rm d}}

\newcommand{\sptlr}{\text{SPT-3G D1}}
\newcommand{\ACTDR}{\textsc{ACT-DR6}} 
\newcommand{\rv}{$r_\mathrm{v}$}
\newcommand*{\planck}{\textit{Planck}}
\newcommand{\vmodes}{$\mathcal{V}$-modes}
\newcommand{\tmodes}{$\mathcal{T}$-modes}
\newcommand{\LCDM}{$\Lambda$CDM}
\newcommand{\allData}{SPA + SPTpolBB + BK18}

\newgeometry{top=1.5cm}

\newcommand{\VM}{\Phi}

\abstract{
We present new constraints on gravitational vector perturbations (\vmodes{}) using Cosmic Microwave Background (CMB) data, including temperature and $E$-mode polarization from \sptlr{}, \ACTDR{}, and \planck{}, as well as $B$-mode data from BICEP/Keck and SPTpol, which provide the strongest constraints on \vmodes{}. We consider three initial conditions (ICs) that source \vmodes{}: neutrino isocurvature (ISO), neutrino octupole (OCT), and a sourced mode (SMD) generated by an anisotropic stress before matter–radiation equality. We also consider including tensor modes along with \vmodes{} for each of these ICs. Combining all datasets, we obtain 95\% confidence level upper limits of $r_\mathrm{v} < 1.3\times10^{-4}$ (ISO), $r_\mathrm{v} < 6.8$ (OCT), and $r_\mathrm{v} < 4.2$ (SMD), with slightly tighter bounds when tensors are included, at a pivot scale $k_p\ =\ 0.05$ Mpc$^{-1}$. Interestingly, for SMD without tensors, using SPTpol $B$-modes alone yields $r_\mathrm{v} = 4.7 \pm 2.1$, consistent with zero at $2.2\sigma$. Similar result is found for SMD when including tensor perturbations. No statistically significant deviation from $\Lambda$CDM is found. However, \vmodes{} are not fully excluded by current $B$-mode data and should be considered when interpreting primordial signals.
}
\begin{document}
	
	\title{The Status of Gravitational Vector Perturbations with Recent CMB Data}

	\author[a]{Ali Rida Khalife,}
	\emailAdd{ridakhal@iap.fr}
	
	\author[a]{Cyril Pitrou}
	\emailAdd{pitrou@iap.fr}
	\affiliation[a]{Institut d'Astrophysique de Paris, CNRS UMR 7095, 98 bis Bd Arago, 75014 Paris, France.}
\date{\today}
\maketitle
\section{Introduction}
\label{Sec:Intro}

The General theory of Relativity (GR) predicts three types of perturbations: scalar, vector and tensor~\cite{Peter-Uzan,Dodelson,Bardeen:1980kt,Kodama_Sasaki,Mukhanov:1990me}. They differ from each other under infinitesimal gauge transformations, and by their impact on cosmological observables, specifically the Cosmic Microwave Background (CMB)~\cite{HuAndWhite,PitrouModes,Hu_Sugiyama,Hu:2000ee}. Most efforts in the last couple of decades have been dedicated to scalar and tensor modes, but not as much so for vector perturbations (\vmodes{}). Historically, the reason for that is the decaying nature of \vmodes{} when assuming standard, adiabatic ICs, which results in a negligible impact on the CMB. 

The decaying nature of adiabatic \vmodes{} implies an irregular behavior when extrapolated backward in time, as the Universe would be in a contracting phase~\cite{Brandenberger_Vmodes}\footnote{Note that this is also a problem for bouncing universe models within GR~\cite{Bouncing1,Bouncing2}}. This irregularity can result in breaking of perturbation theory, which one might consider as a challenge to GR. However, if there are mechanisms in the primordial universe (e.g. primordial magnetic fields~\cite{Kosowsky_etal:2002,Paoletti_theory,Ichiki_etal:2008gr,Takahashi:2007ds,Durrer:1998ya,Kahniashvili:2005xe,Paoletti:2022gsn,Paoletti_Forecasts,Itchiki_etal2006bq,Minoda:2020bod}) that can source the comoving \vmodes{}, then they avoid this issue and can leave an imprint on the CMB. Hence, \vmodes{} are an advantageous window to better understand the primordial universe. 

Recently~\cite{Vmodes_Paper}, we investigated three distinct initial conditions (ICs) for gravitational vector modes (\vmodes{}) to see whether they can be a plausible magnetogenesis mechanism. We found that it is not the case. However, the presence of \vmodes{} is not completely ruled out by the available data. 

In this article, we follow up on our findings in~\cite{Vmodes_Paper} and present constraints on these \vmodes{} ICs using temperature and $E$-mode polarization data from \sptlr{}~\cite{Camphuis_etal,SPT_Maps}, \ACTDR{}~\cite{ACTDR6_main,ACTDR6_Extended,ACTDR6_Maps} and \planck{}~\cite{Planck2018,Planck_Legacy}. We also include $B$-mode polarization data from SPTpol~\cite{SPTpol,SPTpol_candl} and BICEP/Keck~\cite{BICEP_Keck}. As we will see later, $B$-modes are the most impactful in providing strong constraints on the parameters of \vmodes{}.

The article is organized as follows. We give a concise review of the theory behind \vmodes{} in Section~\ref{Sec:Summary}, followed by a description of the data sets and numerical implementation considered in Section~\ref{Sec:Data_Sets_Numerical_Setup}. Our main results, presented in Section~\ref{Sec:Results}, are divided into two parts. Section~\ref{Sec:Res_Vmodes_Only} presents constraints on each of the three ICs of \vmodes{} when they are present along with only scalar perturbations, while in Section~\ref{Sec:Res_Tmodes_Only} we present constraints when also tensor perturbations (\tmodes{}) are present. We finish the article with a brief discussion and conclusion in Section~\ref{Sec:Conclusion}.

\section{Review of \vmodes{} and their Initial Conditions}
\label{Sec:Summary}
In this section, we make a short summary of the theory of \vmodes{} and of the three ICs we consider (for more details, see~\cite{Vmodes_Paper} and references therein). Recall that, at linear order, the three types of perturbations evolve independently. However, they all contribute additively to the CMB spectra, as we shall see shortly.
\subsection{\vmodes{}}
We consider a spatially flat Friedmann-Lema\^itre-Robertson-Walker (FLRW) metric of the form
\be 
g_{\mu\nu} = a^2(\eta)\big(\eta_{\mu\nu}+h_{\mu\nu}\big),
\label{Eq:Metric}
\ee
where $\eta_{\mu\nu}=\mathrm{diag}[-1,1,1,1]$ (the Minkowski metric), $h_{\mu\nu}$ is the metric perturbation and $a(\eta)$ is the scale factor as a function of conformal time, $\eta$. Focusing only on \vmodes{}, we choose the gauge such that the perturbed metric takes the form
\be
h_{00}=0; \ h_{0i} = -\Phi_i; \ h_{ij}=0, 
\label{Eq:Pert_Metric}
\ee
where $\Phi_i$ are the (divergenceless) \vmodes{}. After expanding all the relevant quantities using the normal expansion technique (see Section 2.2 of~\cite{Vmodes_Paper}), we get the following equations of motion for \vmodes{}:
\be
\Phi^{(m)}{}' + 2{\cal H}\Phi^{(m)} = -\frac{8\pi G a^2}{k}\sum_{s} \bar{p}_s \pi_s^{(m)},
\label{Eq:Evol_Phi_ModeFunction}
\ee

\be
\Phi^{(m)} = \frac{16\pi Ga^2}{k^2}\sum_{s}(\bar{\rho}_s+\bar{p}_s) \left(v_s^{(m)}-\Phi^{(m)}\right).
\label{Eq:Constraint_Eq}
\ee
where $\Phi^{(m)}$ (with $m=\pm \ 1$) are the normal modes of $\Phi_i$, $'$ denotes a derivative with respect to (w.r.t) $\eta$, $\mathcal{H}={a}'/a$, $G$ is Newton's constant, $k$ is the wavenumber, $\bar{\rho}_s$ ($\bar{p}_s$) is the background energy density (pressure) of species $s$, $\pi_s^{(m)}$ is its dimensionless anisotropic stress and $v_s^{(m)}$ is its (divergenceless) velocity.

The presence of \vmodes{} will alter the evolution equations of CMB anisotropies. For instance, the evolution equation of the CMB temperature anisotropies takes the form
\be
(\Theta_{\ell}^{(m)}) {}' = k\left[\frac{{}_0 \kappa_{\ell}^m}{2{\ell}-1}
	\Theta_{{\ell}-1}^{(m)} - \frac{{}_0 \kappa_{{\ell}+1}^m}{2{\ell}+3}
	\Theta_{{\ell}+1}^{(m)} \right] +{}^\Theta {\cal C}_{\ell}^m -\tau' \Theta_{\ell}^{(m)} \,, \ (\ell\geq m)\,,
\ee
where $\Theta_{\ell}^{(m)}$ are the temperature multipoles in the Boltzmann hierarchy, $\tau$ is the optical depth,
 \be\label{Defkappaslm}
 {}_s \kappa_\ell^m \equiv\sqrt{\frac{(\ell^2-m^2)(\ell^2-s^2)}{\ell^2}},
 \ee
 and
 \be
  {}^{\Theta}{\cal C}_{\ell}^m = \left(\tau' v_b^{(m)}+\Phi^{(m)}{}'\right)\delta_{\ell1}+\tau'P^{(m)}\delta_{\ell2}; \  P^{(m)} \equiv \frac{1}{10}\left(\Theta_2^{(m)} - \sqrt{6}E_2^{(m)}\right).
 \ee
 Here, $v_b^{(m)}$ is the normal mode of the baryon velocity, $\delta_{ij}$ the Kronecker delta and $E_2^{(m)}$ is the quadrupole ($\ell=2$) of the $E$-mode polarization of the CMB. Moreover, the angular power spectra of the CMB will now have an additional contribution due to the presence of \vmodes{} of the form\footnote{There will be a similar expression for scalar and tensor perturbations. Eventually, the total CMB spectrum is the sum of the three spectra from each type of perturbations.}
 \be
C_{\ell}^{XY} = \frac{2}{\pi(2\ell+1)^2} \int \dd k k^2{\cal T}_X(k,\eta){\cal T}{}^\star_Y(k,\eta) P_\VM(k)
\label{Eq:CMB_Spect}
\ee
where $X,Y=\{\Theta,E,B\}$, ${\cal T}_{X}(k,\eta)$ is the corresponding transfer function (${\cal T}_{X}{}^\star$ its complex conjugate) and
\be
P_{\Phi}=r_\mathrm{v} A_s \bigg(\frac{k}{k_*}\bigg)^{n_v}
\label{Eq:Primodial_V_PS}
\ee
is the primordial \vmodes{} power spectrum. Here, $r_\mathrm{v}=A_\mathrm{v}/A_s$ is the amplitude of the \vmodes{}' power spectrum ($A_\mathrm{v}$) relative to that of scalars ($A_s$), $n_v$ is its spectral index and $k_*\ =\ 0.05$ Mpc$^{-1}$ is a pivot scale. Note that this is the same expression as for the \tmodes{} power spectrum except that, assuming a slowly rolling single field inflation, the spectral index ($n_t$) and amplitude of \tmodes{}' power spectrum are related by the self consistency condition (scc)~\cite{SCC}:
\be
n_\mathrm{t,scc} = -\frac{r}{8}\left(2-\frac{r}{8}-n_s\right),
\label{eq:SCC}
\ee
where $r=A_T/A_s$ is the amplitude of the \tmodes{}' power spectrum ($A_T$) relative to $A_s$, and $n_s$ is the spectral index of the scalar perturbations' power spectrum.


\subsection{The Three Initial Conditions}

We now give a qualitative description of the neutrino isocurvature (ISO)~\cite{Lewis:2004kg,Itchiki_V-modes,Paoletti_theory,Rebhan:1991sr,Rebhan:1994zw}, neutrino octupole (OCT) and sourced mode (SMD) initial conditions (see Section 3 of~\cite{Vmodes_Paper} for more details). The mechanism in common between these three ICs is the need for an initial anisotropy ($\pi^{(m)}_s$) to source the \vmodes{}, otherwise they decay as $a^{-2}$ and will not survive long enough to make an impact on the CMB (see eq.~\eqref{Eq:Evol_Phi_ModeFunction}). 

\begin{itemize}
    \item \textbf{ISO}: Consider the situation where the early Universe is described by a theory beyond the Standard Model (SM) of particle physics. In this case, in order to source the \vmodes{} through eq.~\eqref{Eq:Evol_Phi_ModeFunction}, one must have modifications to the weak nuclear interactions such that neutrinos could decouple with a velocity different from that of photons. This velocity difference will create a neutrino anisotropic stress ($\pi^{(m)}_{\nu}$) that sources the \vmodes{}.  

    \item \textbf{OCT}: In the primordial Universe, all particles are in equilibrium and tightly coupled to each other. This means that all particles, specifically photons and neutrinos, have their Boltzmann hierarchy truncated at the quadrupole ($\ell=2$) level. However, if there are mechanisms beyond the SM that can break this tight coupling, this would be compatible with a non-vanishing value of the higher multipoles, in particular the neutrino octupole ($\ell=3$). In this case, the neutrino octupole will source its quadrupole, which will then source the \vmodes{}, allowing them (in principle) to survive long enough to have an imprint on the CMB.

    \item \textbf{SMD}: In this IC, we consider a toy model in which the \vmodes{} are directly sourced, at a redshift $z_\mathrm{start}$ by an anisotropic stress of agnostic origin (this could be from topological defects~\cite{PMF_Defects4}, for instance). More specifically, we consider an anisotropic stress of the form:
        \be
            \pi_s= \pi_*\delta(z-z_\mathrm{start}),
            \label{Eq:DS_AnisotropicStress}
        \ee
    where $\pi_*$ is a constant and $\delta(z-z_\mathrm{start})$ is the Dirac delta function. We then compute the amplitude of \vmodes{} relative to that at an arbitrary reference redshift, which we set to be $z_\mathrm{ref}=10^4$. Although this IC is more phenomenological than the previous two, it has a better chance of being explained by additional primordial physics, which is a topic that we will consider in a future work. 
\end{itemize}

\begin{figure}[!htb]
		\begin{subfigure}[t]{0.57\textwidth}
			\hspace{-1.9cm}
			\centering
			\includegraphics[width=\textwidth]{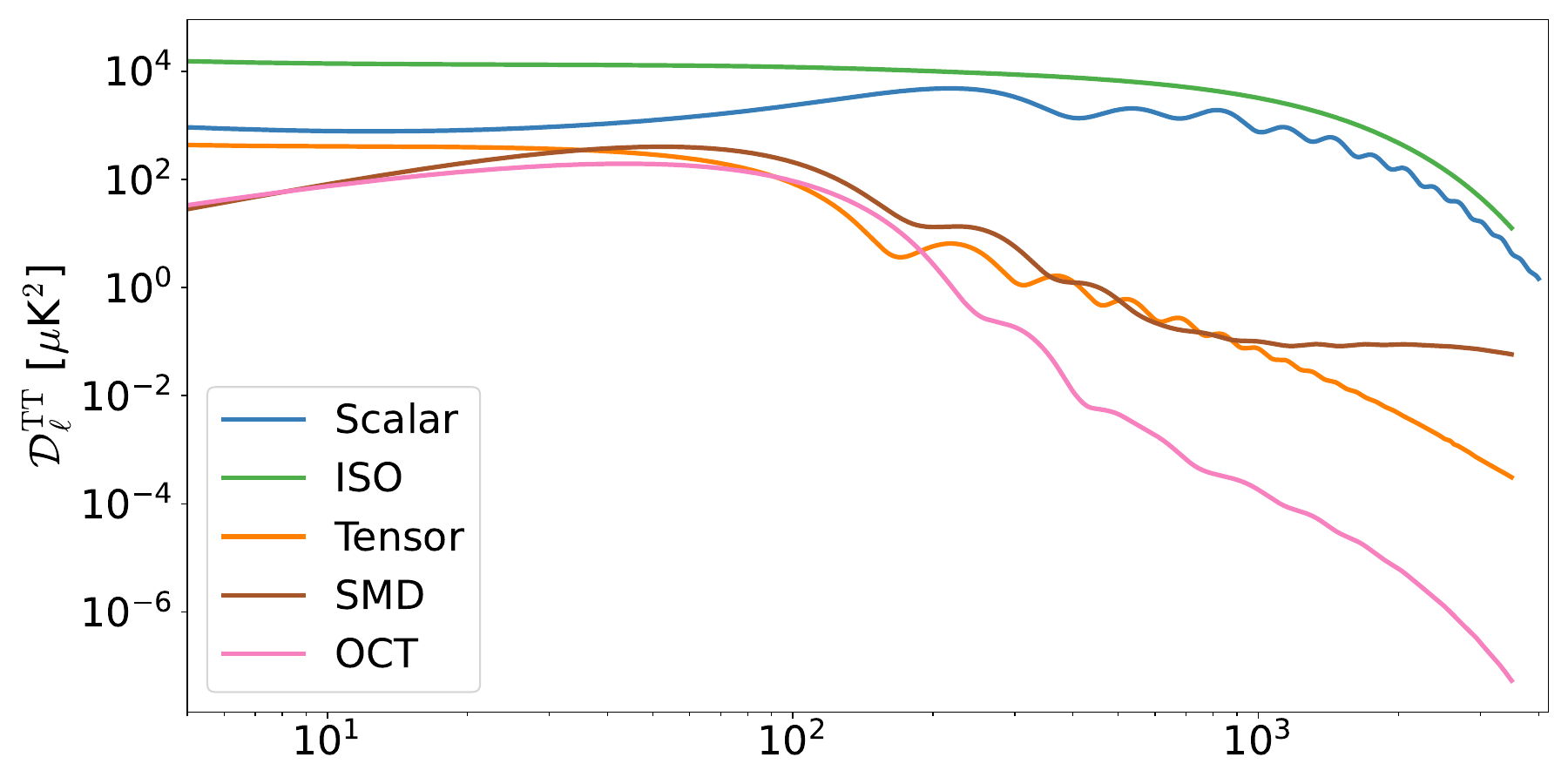}
		\end{subfigure}%
		\hfill
		\begin{subfigure}[t]{0.57\textwidth}
			\hspace{-1.9cm}
			\centering
			\includegraphics[width=\textwidth]{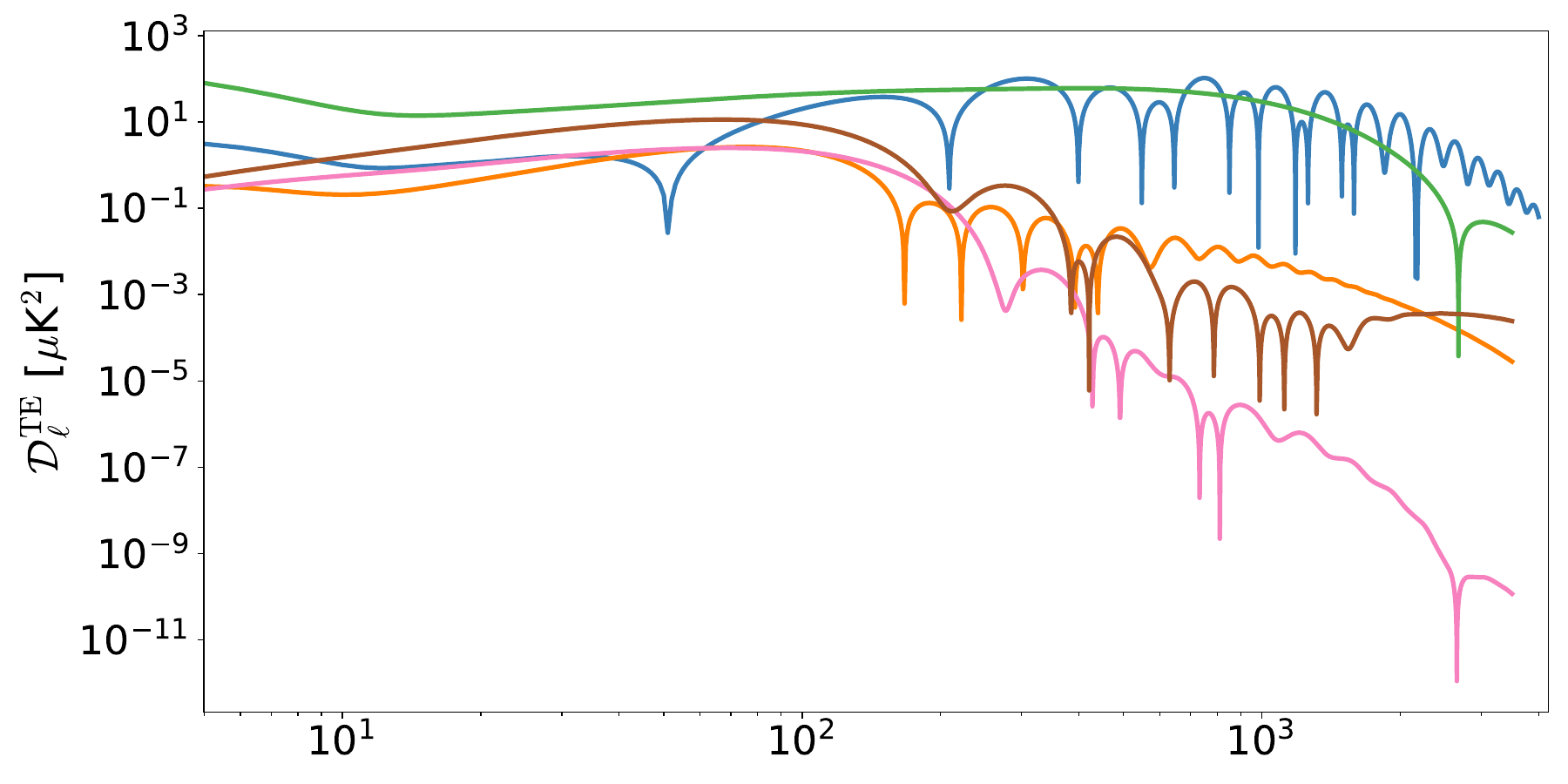}
		\end{subfigure}
		\vspace{0.5cm}
		\begin{subfigure}[t]{0.57\textwidth}
			\hspace{-1.9cm}
			\centering
			\includegraphics[width=\textwidth]{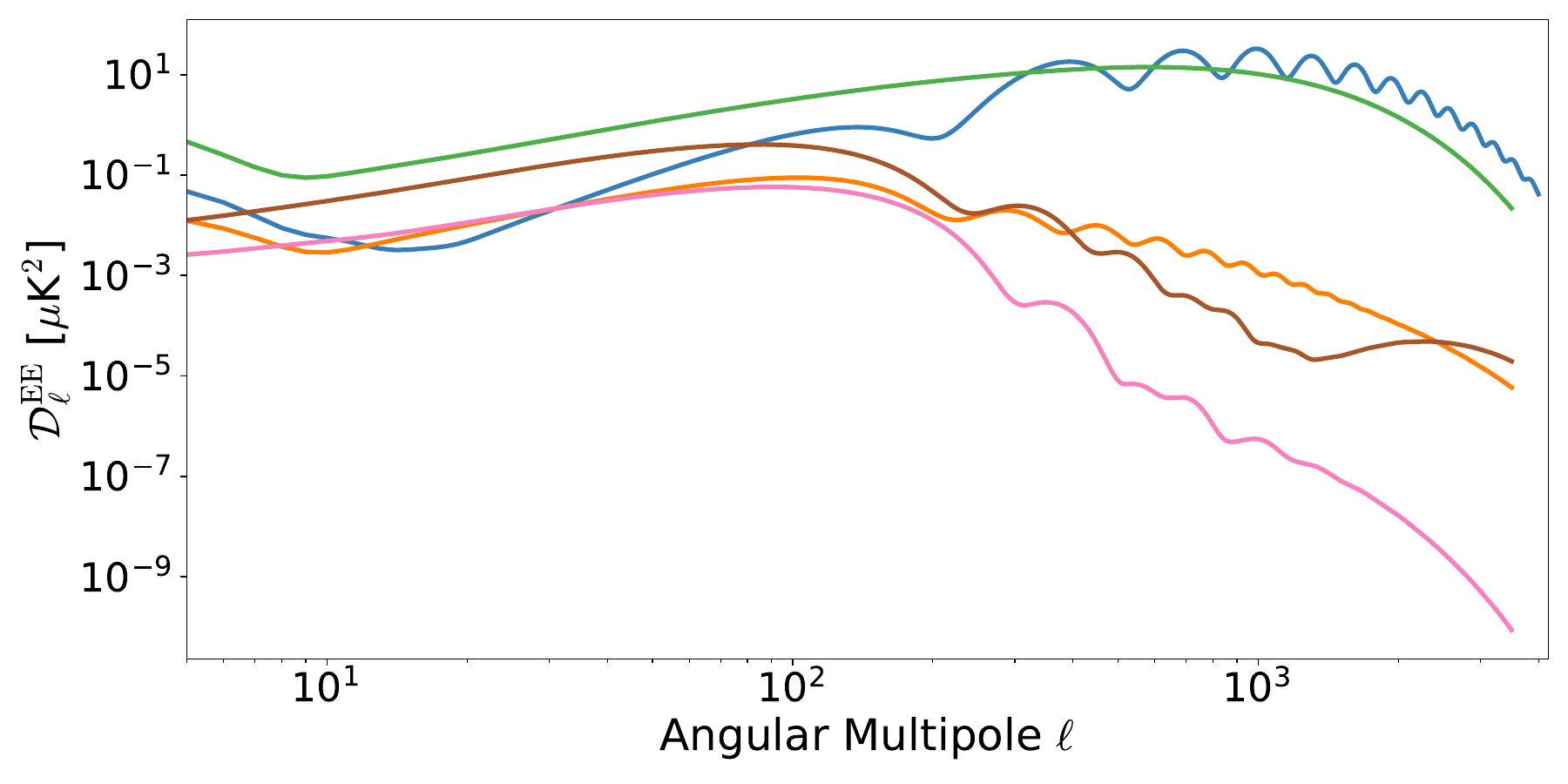}
		\end{subfigure}%
		\hfill
		\begin{subfigure}[t]{0.57\textwidth}
			\hspace{-1.9cm}
			\centering
			\includegraphics[width=\textwidth]{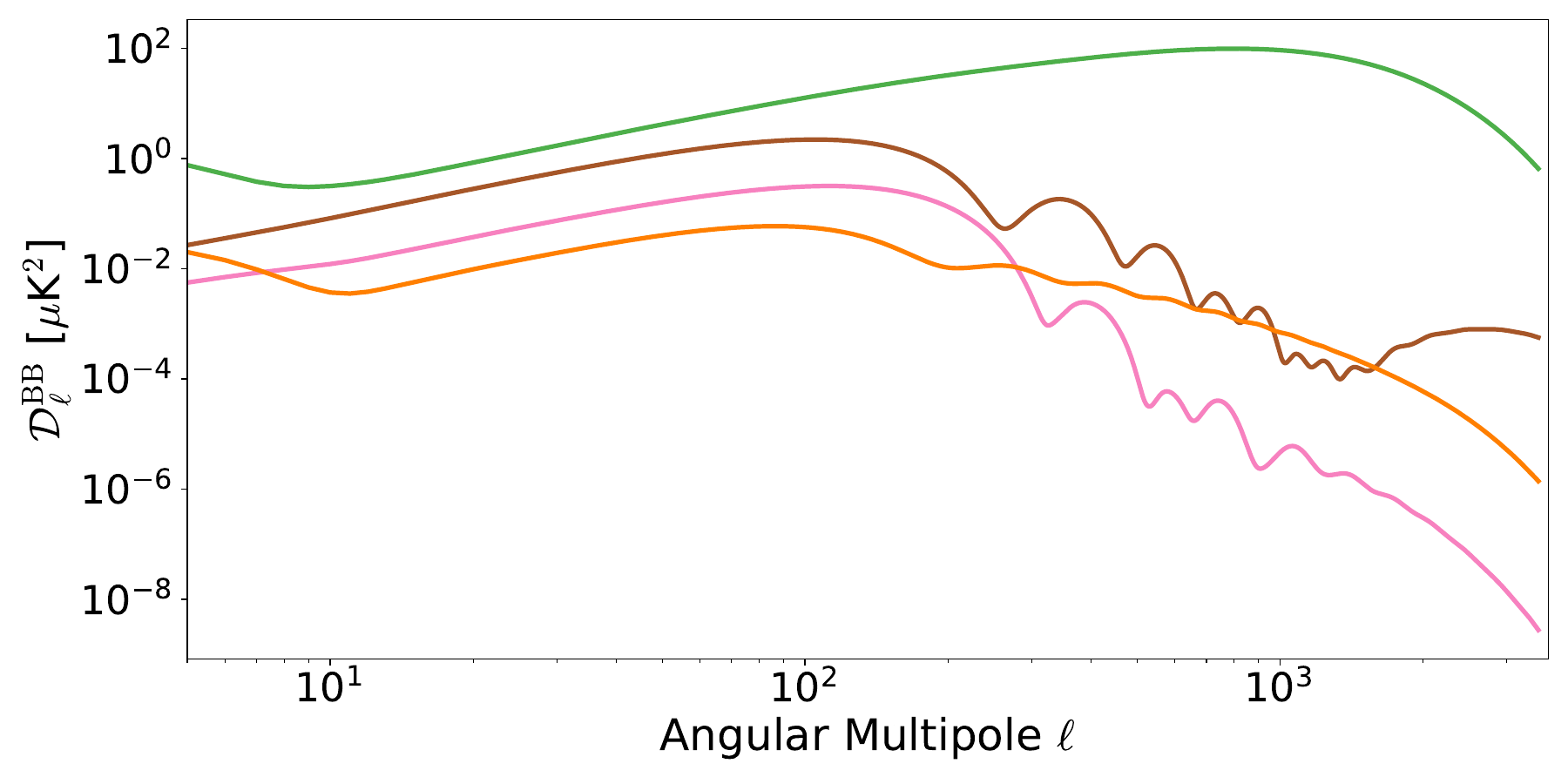}
		\end{subfigure}
		\caption{\justifying CMB spectra, ${\cal D}_{\ell}$ (eq.~\eqref{Eq:D_ell}), for the three different $\mathcal{V}$-modes considered in this work compared to those of Scalar and Tensor modes. For illustration, we set the $\Lambda$CDM cosmological parameters to those given by the \planck{} best-fit~\cite{Planck_Primary},
        $r_\mathrm{v}=r=1$ and $n_v=n_t=n_s$. For the SMD case, we set $z_*=10^7$. This figure is a replication of Figure 10 in~\cite{Vmodes_Paper}.}
\label{Fig:CMB_Spectra}
\end{figure}

We finish this section by showing the shape of the CMB spectra,
\be
{\cal D}_{\ell}^{XY} = \frac{\ell(\ell+1)}{2\pi}C_{\ell}^{XY}
\label{Eq:D_ell}
\ee
due to the different types of perturbations in Figure~\ref{Fig:CMB_Spectra}. We set the amplitudes and spectral indices of all the perturbations equal to each other in order to focus on the intrinsic difference between these spectra.

\section{Data Sets and Numerical Setup}
\label{Sec:Data_Sets_Numerical_Setup}
In this work, we consider the following CMB data sets (abbreviated by the name in bold font) in constraining \vmodes{}:

\begin{itemize}

    \item \textbf{SPA}: Temperature ($TT$), $E$-mode polarization ($EE$) and their cross-correlation ($TE$) data from \sptlr{}~\cite{Camphuis_etal,SPT_Maps}, \ACTDR{}~\cite{ACTDR6_Maps,ACTDR6_main,ACTDR6_Extended} and \planck{}~\cite{Planck2018,Planck_Legacy}.\footnote{Note that we do not consider any lensing reconstruction data from the CMB in this work.} We replace the low-$\ell\ EE$ information from \planck{} with a gaussian prior on the optical depth to reionization, $\tau_\mathrm{reio}$, of the form $\mathcal{N}\left(0.051,0.006^2\right)$~\cite{planck_collaboration_planck_2020}. To avoid double counting information when combining \ACTDR{} and \planck{}, we cut the \planck{} $TT$ spectrum at $\ell>1000$, while $\ell>600$ is cut for the $TE$ and $EE$ spectra.
    
    \item \textbf{SPTpolBB}: $B$-mode polarization ($BB$) data from the SPTpol survey~\cite{SPTpol}. We use the \texttt{candl}~\cite{candl} implementation of this likelihood, as done in~\cite{SPTpol_candl}.

    \item \textbf{BK18}: $BB$ data from BICEP/Keck XIII survey~\cite{BICEP_Keck}. When this dataset is combined with SPTpolBB, we restrict the latter to the multipole range $\ell>400$ to avoid double counting of information.\footnote{Cutting the SPTpol data when combining with BK18 is not strictly necessary, given the higher noise of the SPTpol $BB$ measurements compared to BK18 in the overlapping $\ell$-range. Nevertheless, we adopt this cut as a conservative choice to avoid any potential double counting.}
    
\end{itemize}

Note that, except for BK18, we use the \texttt{lite}, i.e. foreground-marginalised~\cite{CMB_lite,Lennart_lite} versions of the likelihoods.

For the numerical implementation, we modified the Boltzmann code \texttt{CLASS}~\cite{CLASSI,CLASSII} to incorporate  \vmodes{} for each of the above mentioned ICs and perform our computations (which we make publicly available\footnote{\url{https://github.com/CyrilPitrou/class/tree/class_vectors_B}}). We use the Bayesian analysis code \texttt{COBAYA}~\cite{COBAYA1,COBAYA2,COBAYA3}\footnote{\url{https://cobaya.readthedocs.io/en/latest/}} to run our Markov chains Monte Carlo (MCMC), and consider chains to be converged when the Gelman-Rubin criterion~\cite{Gelman_Rubin} is $R-1\lesssim 0.05$. In addition to $\tau_\mathrm{reio}$ mentioned above, we sample over the energy densities of baryons $\left(\omega_b\right)$ and cold dark matter $\left(\omega_b\right)$, the angular size of the sound horizon at recombination $\left(\theta_s\right)$, $n_s$, $\ln\left(10^{10}A_s\right)$, \rv{}, and $n_v$. We additionally sample over $r$ in the case of adding \tmodes{}, while $n_t$ is fixed to the scc eq.~\eqref{eq:SCC}. We impose uniform priors on all these parameters. To find the best-fit parameters and $\chi^2$s for the \allData{} data combination, we use the \texttt{BOBYQA} algorithm~\cite{BOBYQA_I,BOBYQA_II}\footnote{\url{https://github.com/numericalalgorithmsgroup/pybobyqa}} (as implemented in \texttt{COBAYA}), except for the SMD IC (both with and without \tmodes{}) and for the OCT IC in the case where both \tmodes{} and \vmodes{} are included due to numerical instabilities encountered with \texttt{BOBYQA}. In this case, we employ the code \texttt{PROSPECT}~\cite{Prospect}\footnote{\url{https://github.com/AarhusCosmology/prospect_public}} to determine the best fit.


\section{Results}
\label{Sec:Results}

We present our results by first considering the scenario where \vmodes{} are present alone (in addition to scalar perturbations), and then a second scenarios corresponding to the case where scalar, vector and tensor perturbations co-exist. Constraints of the first scenario are concisely presented in Table~\ref{Tab:Vmodes_only} and Figure~\ref{fig:Vmodes_only}, while Table~\ref{Tab:Vmodes_Tmodes} and Figure~\ref{fig:Vmodes_Tmodes} correspond to the second scenario. Throughout the rest of the article, constraints on \rv{} or $r$ correspond to upper limits at 95\% confidence level (C.L.), in addition to a 68\% confidence region (when applicable). For the other parameters, we show the 68\% confidence region.

\subsection{Vector Modes Alone}
\label{Sec:Res_Vmodes_Only}
Before discussing specific results for each IC, we note a general trend in the constraints on \vmodes{}. First, CMB $B$-modes are crucial to greatly improve the constraints on \rv{}. From SPA alone, the upper limit on \rv{} is almost 1 order of magnitude larger than those with SPTpolBB or BK18 (except for the OCT case, as we will note shortly). This is yet another motivation for having high precision $B$-modes measurements. However, $n_v$ is well constrained already from SPA alone. The reason is that changing the tilt will impact the small scales of $TT$, $TE$ and $EE$ spectra, where \sptlr{} and \ACTDR{} data are well constraining. This does not allow for a lot of leeway in $n_v$. 

\begin{figure}[!htp]
    \centering
    \hspace{-3cm}
    \begin{subfigure}{0.45\textwidth}
        \centering
        \includegraphics[width=1.3\linewidth]{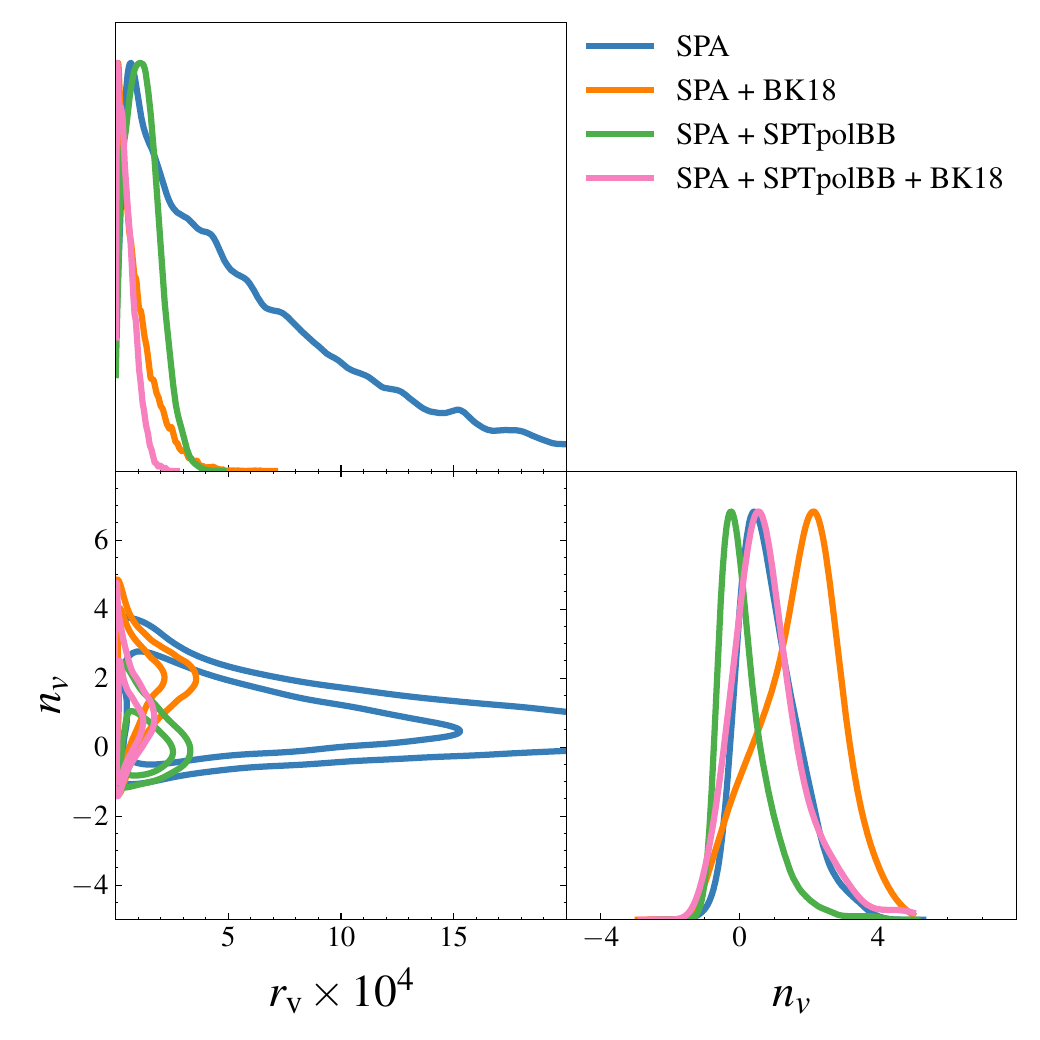}
        
        \caption{ISO}
        \label{fig:Vmodes_only_ISO}
    \end{subfigure}
    \hfill
    \begin{subfigure}{0.45\textwidth}
        \centering
        \includegraphics[width=1.3\linewidth]{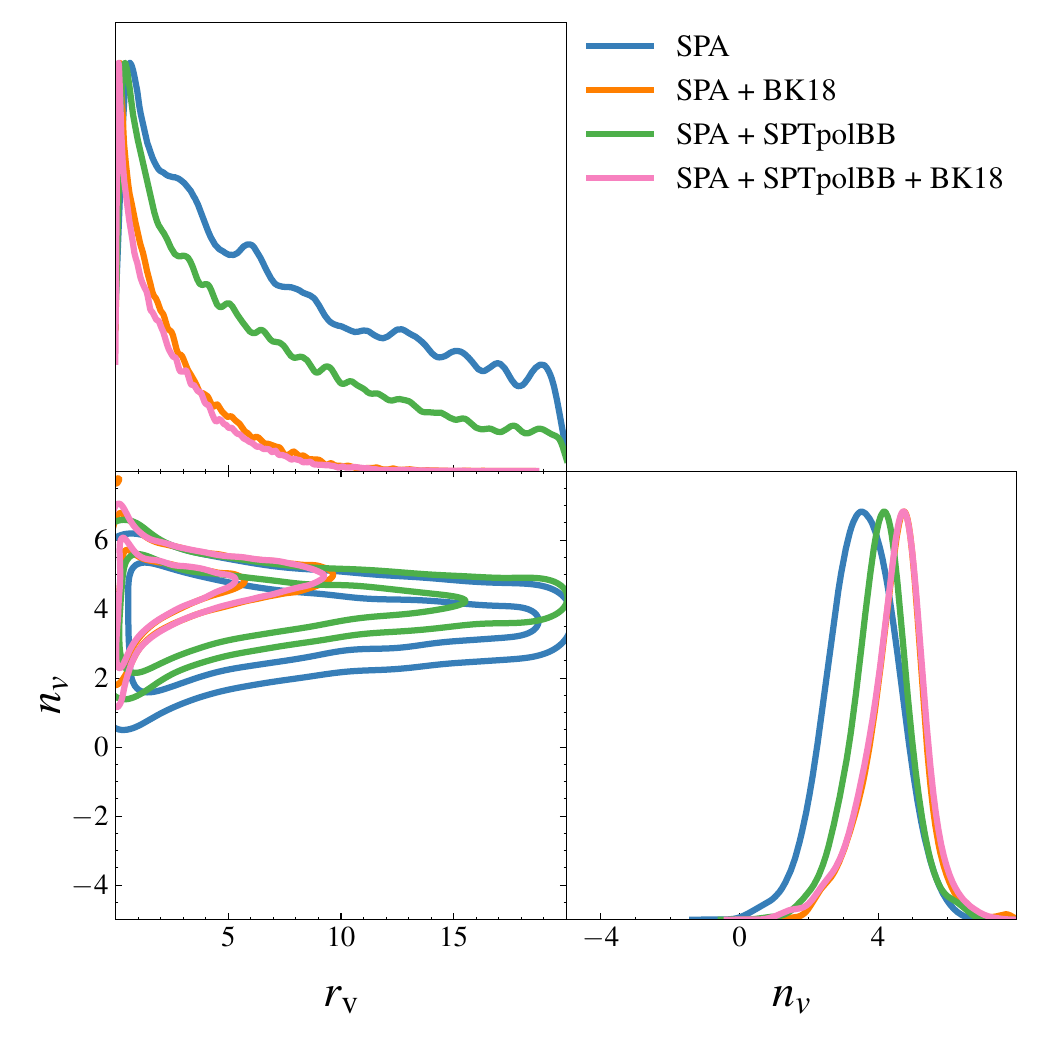}
        \caption{OCT}
        \label{fig:Vmodes_only_OCT}
    \end{subfigure}
    
    \vspace{0.5cm} 
    
    \hspace{-3cm}
    \begin{subfigure}{0.45\textwidth}
        \centering
        \includegraphics[width=1.3\linewidth]{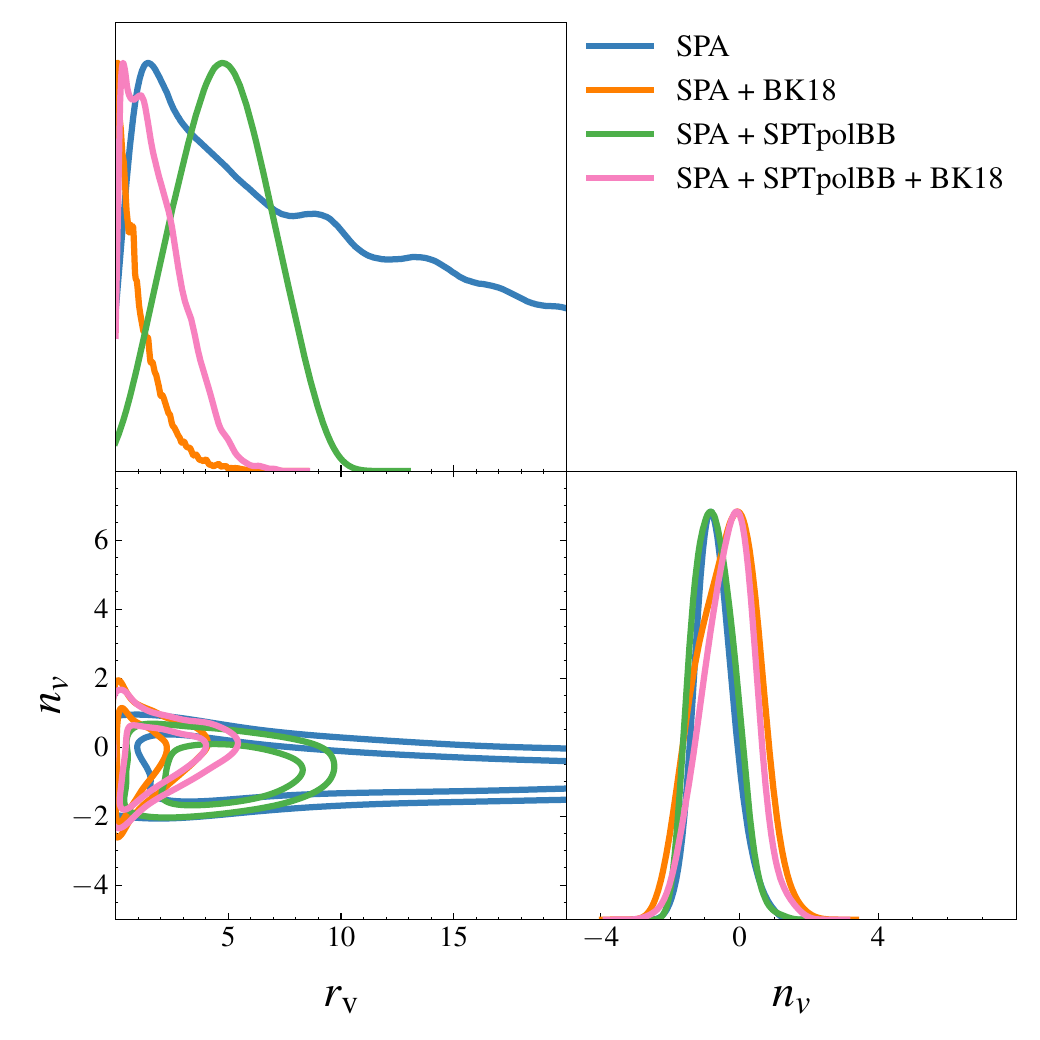}
        \caption{SMD $\left(z_\mathrm{start}=10^4\right)$}
        \label{fig:Vmodes_only_SMD_fix_zs}
    \end{subfigure}
    \hfill
    \begin{subfigure}{0.45\textwidth}
        \centering
        \includegraphics[width=1.3\linewidth]{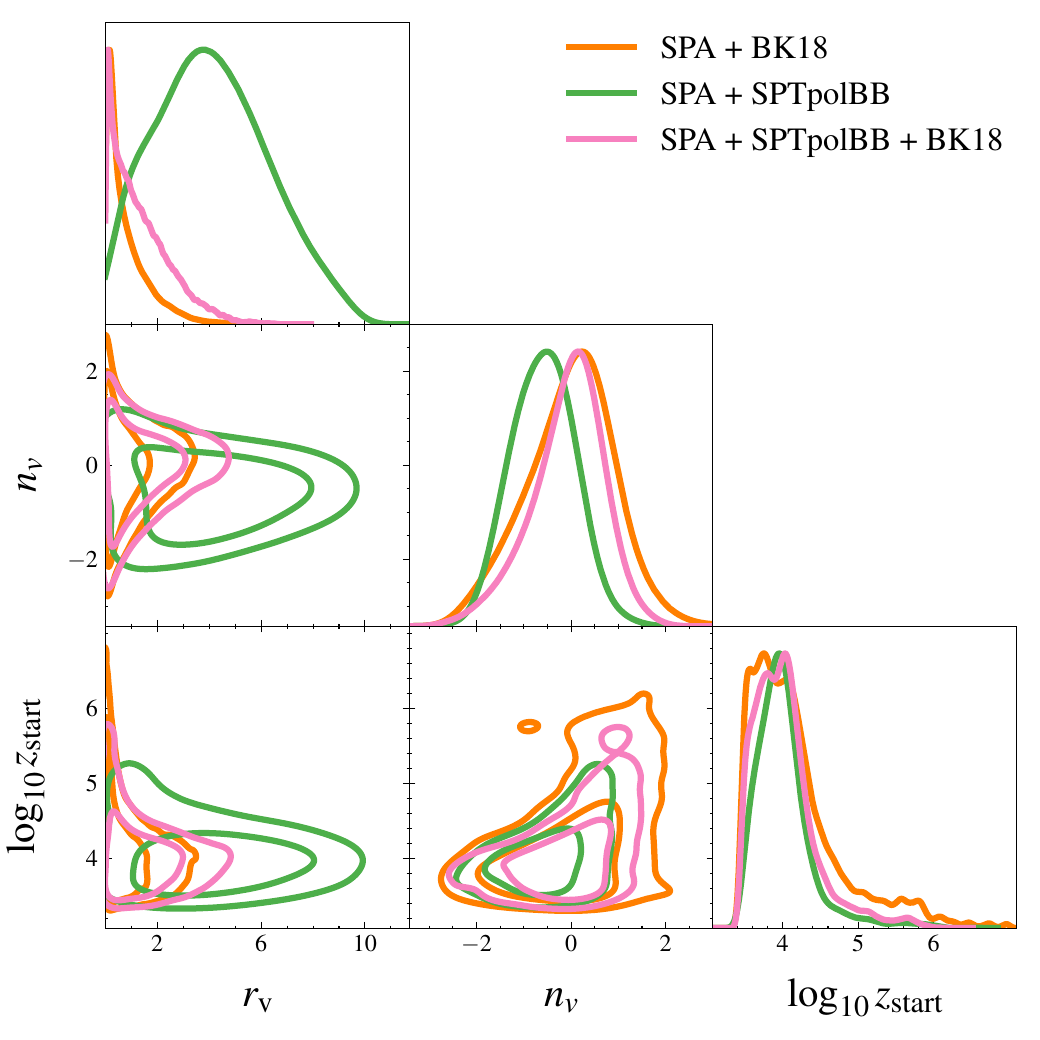}
        \caption{SMD (varying $z_\mathrm{start}$)}
        \label{fig:Vmodes_only_SMD_vary_zs}
    \end{subfigure}

    \caption{Constraints on the amplitude of \vmodes{} power spectrum relative to scalar perturbations (\rv) and its spectral index ($n_v$) for ISO (a), OCT (b) and SMD (c) ICs from different data combinations, as indicated in the legends (see also Section~\ref{Sec:Data_Sets_Numerical_Setup}). We also consider the case where, for the SMD IC, the redshift at which \vmodes{} start being sourced ($z_\mathrm{start}$) is a free parameter (d).}
    \label{fig:Vmodes_only}
\end{figure}

Starting with the ISO case, we can see from Table~\ref{Tab:Vmodes_only} and Figure~\ref{fig:Vmodes_only} that the \rv{} constraint here is $4-5$ orders of magnitude smaller than in the other two ICs. \vmodes{} sourced by an ISO IC change the photon dipole directly, which impacts the CMB spectra more strongly compared to the other cases (see Figure~\ref{Fig:CMB_Spectra}), resulting in a tighter constraint. Already with SPA we get $r_\mathrm{v}\ <\ 19\times10^{-4}$, and then adding SPTpolBB and BK18 results in the most up to date upper limit of $r_\mathrm{v}\ <\ 1.3\times10^{-4}$ for \vmodes{} with the ISO IC\footnote{Although our constraints are not exactly comparable to those in~\cite{Saga:2014zra} due to the difference in pivot scale, the comparison is approximately valid and shows an improvement in constraints on \rv{} by almost a factor of 6.}. Concerning $n_v$, the ISO IC shows a slightly blue-tilted (but still consistent with a scale invariant) power spectrum, with $n_v\ =\ 0.81^{+0.69}_{-1.1}$ using the \allData{} data combination. However, we do not find any statistical preference for this model compared to \LCDM{} with $r_\mathrm{v}=0$, since we find a $\Delta\chi^2\ =\ \chi^2_\mathrm{model} - \chi^2_{\Lambda\mathrm{CDM}} \ =\ -1.2$.

For the OCT case, whose constraints are shown in the second block of Table~\ref{Tab:Vmodes_only} and in Figure~\ref{fig:Vmodes_only_OCT}, we highlight two features. First, we find a comparable upper limit with SPA ($r_\mathrm{v}\ <\ 18$) and from SPA + SPTpolBB ($r_\mathrm{v}\ < \ 17$), the latter being the largest among all ICs. However, when we consider BK18 data, the upper limit reduces by more than half, where now we have $r_\mathrm{v}\ <\ 7.2$ and $r_\mathrm{v}\ <\ 6.8$ for SPA + BK18 and \allData{}, respectively. Second, compared to the other ICs, the OCT case has a highly blue-tilted spectrum of \vmodes{}, with $n_v\ =\ 4.47^{+0.93}_{-0.66}$ for \allData{}. Note that, as in the previous IC, we still do not find any statistical preference over \LCDM{} with \rv{} = 0 ($\Delta\chi^2\ =\ -0.56$).

\begin{figure}[!htbp]
	\centering
	\includegraphics[width=1.\textwidth]{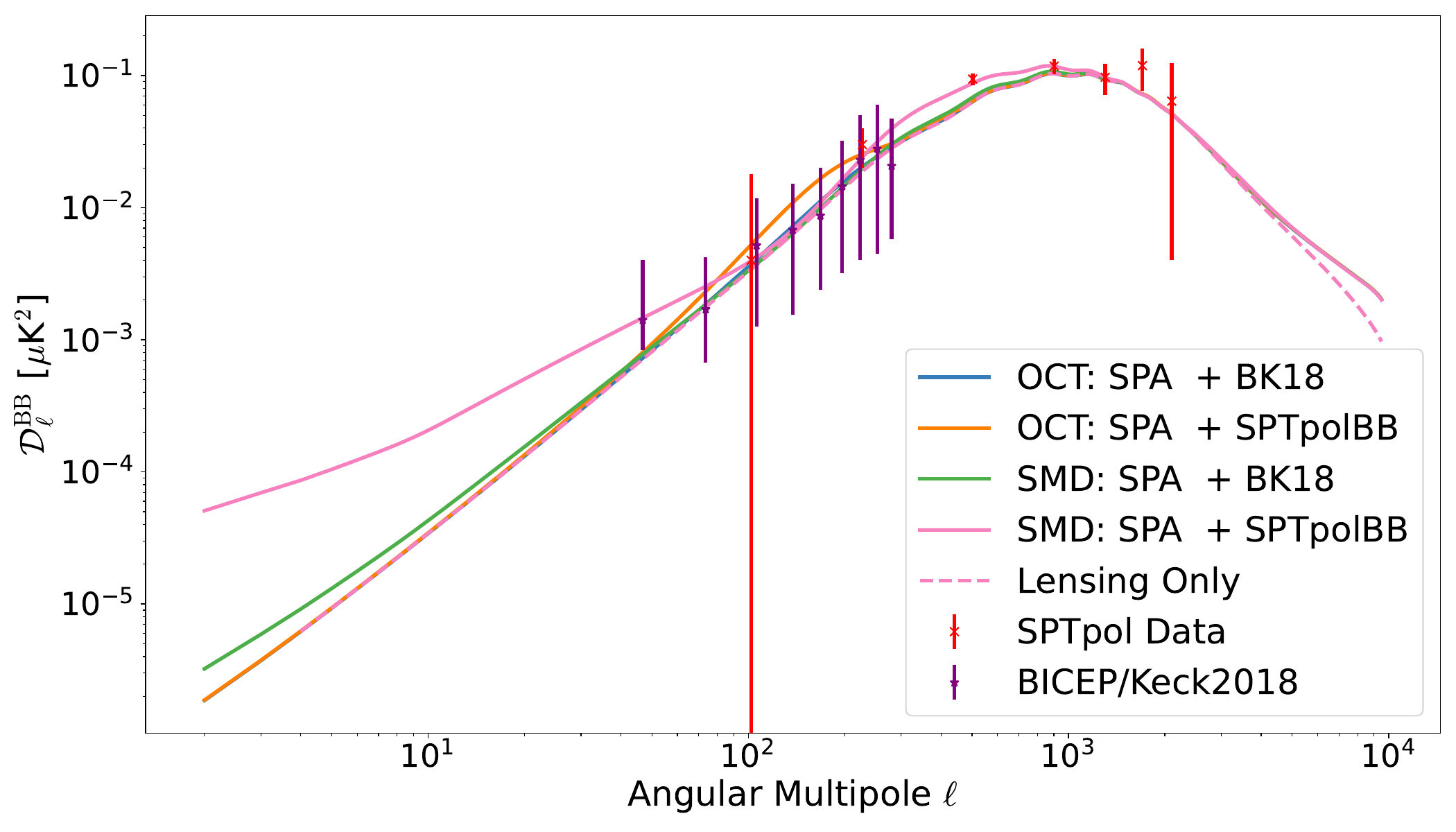}
	\caption{Total CMB $B$-mode power spectra for the OCT IC constrained with SPA + BK18 (SPA + SPTpolBB) in blue (orange), and for the SMD IC constrained with SPA + SPTpolBB (SPA + BK18) in green (pink). As reference, we also show the $B$-modes spectrum due to lensing in dashed pink, which is common to all ICs. Each spectrum is produced from the mean values of the cosmological parameters for the corresponding case. Also shown are the $B$-modes bandpowers from BK18~\cite{BICEP_Keck} in purple and SPTpolBB~\cite{SPTpol} in red. Among all cases, only the SMD IC constrained with SPA + SPTpolBB yields a spectrum that passes through the SPTpolBB bandpowers at $\ell \sim 500$ and $\ell \sim 900$, driving the peak observed in the $r_\mathrm{v}$ posterior for this scenario (see Figures~\ref{fig:Vmodes_only_SMD_fix_zs},~\ref{fig:Vmodes_only_SMD_vary_zs}, and~\ref{fig:Vmodes_Tmodes_SMD}).}
    \label{fig:BB_spec_SMD}
\end{figure}

\begin{figure}[htbp]
\centering

\begin{subfigure}{0.45\textwidth}
\hspace{-3cm}
    \includegraphics[width=1.4\linewidth]{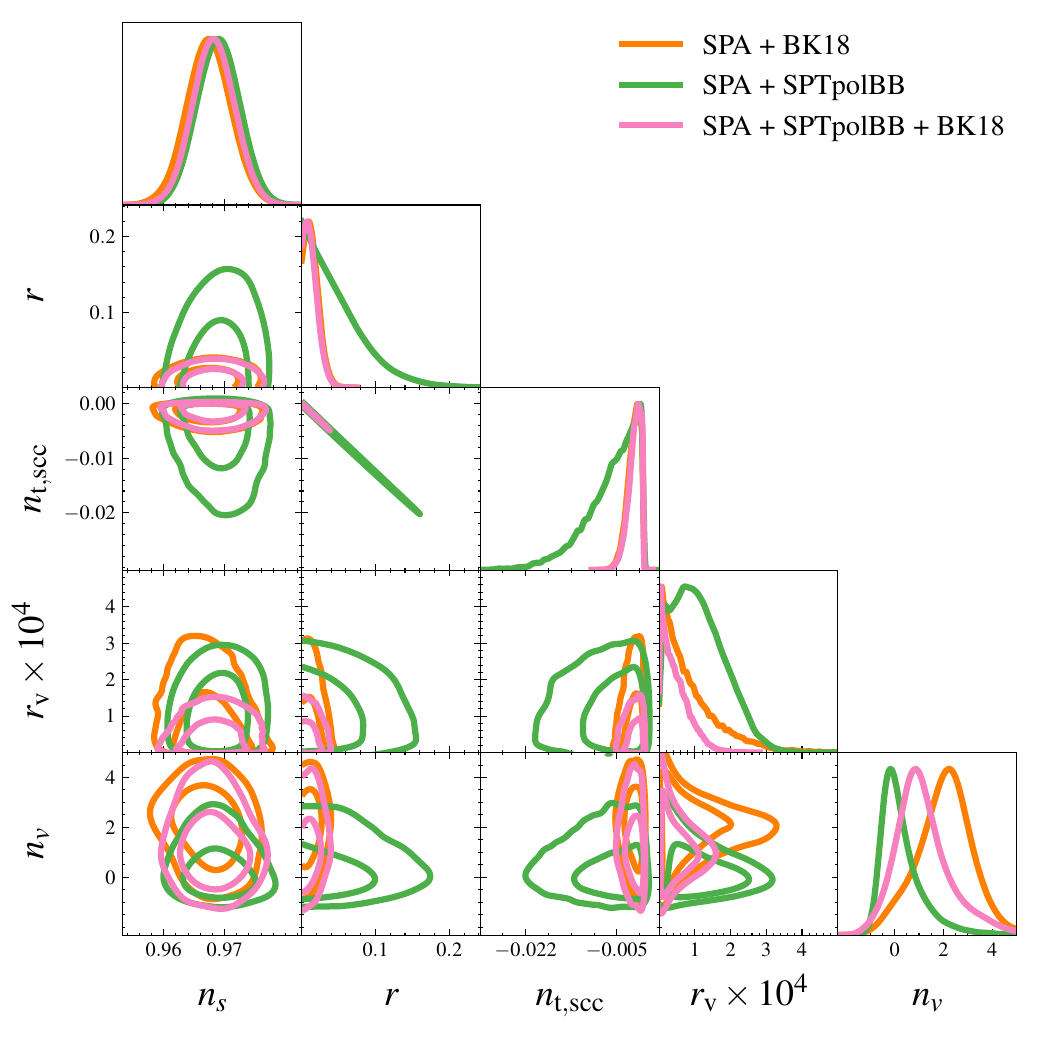}
    \caption{ISO}
\end{subfigure}
\hfill
\begin{subfigure}{0.45\textwidth}
    \hspace{-1cm}
    \includegraphics[width=1.4\linewidth]{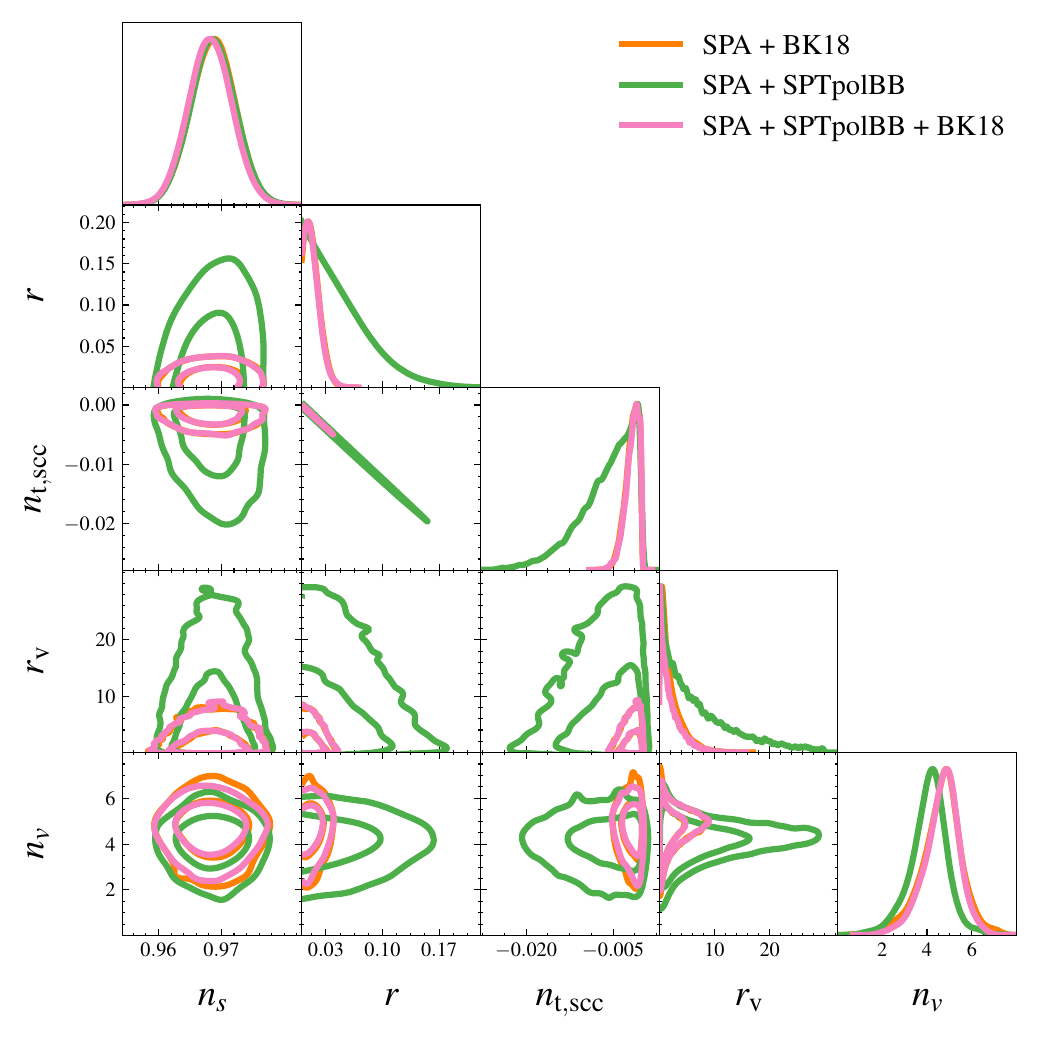}
    \caption{OCT}
\end{subfigure}

\vspace{0.5cm}

\begin{subfigure}{0.45\textwidth}
    \centering
    \includegraphics[width=1.4\linewidth]{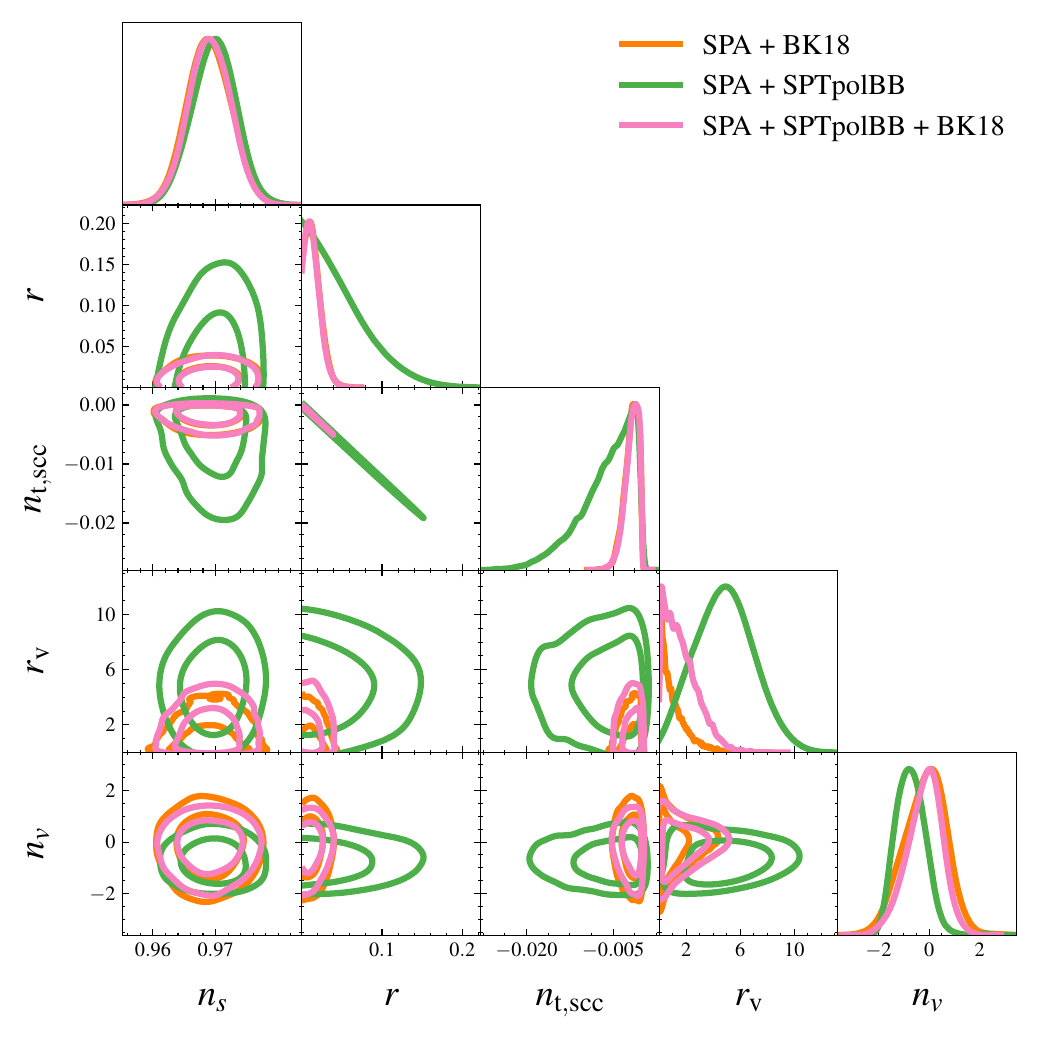}
    \caption{SMD}
    \label{fig:Vmodes_Tmodes_SMD}
\end{subfigure}

\caption{Same as Fig.~\ref{fig:Vmodes_only} but including \tmodes{}.}
\label{fig:Vmodes_Tmodes}

\end{figure}

The origin of these features can be explained by recalling that the OCT IC has the least direct impact on CMB photons; a non-vanishing neutrino octupole sources the quadrupole, which sources the \vmodes{} which then alter the CMB spectra. To see this more clearly, we can look at the OCT spectra in Figure~\ref{Fig:CMB_Spectra}. The CMB spectra for this IC start decaying quickly already at $\ell \sim 150$. To lift the spectra to where the data points are, we need larger power at smaller scales, i.e. a blue-tilted spectrum. This is also related to the fact that, on small scales, the \vmodes{} decay much earlier and quicker in this IC compared to the ISO one (see Figure 4 of~\cite{Vmodes_Paper}). Even with this tilt, a small \rv{} would not be sufficient to match the data points of SPTpolBB (see Figure~\ref{fig:BB_spec_SMD}), which eventually results in a larger upper limit when considering this data set. However, on large scales ($\ell\ \lesssim\ 100$), Figure~\ref{Fig:CMB_Spectra} shows that the OCT spectra are comparable to the other cases, specially the $E$-modes one which will be converted to $B$-modes due to lensing. Thus, in order not to overproduce $B$-mode power and match the BK18 data, \rv{} should decrease.

Finally, for the SMD IC, we consider two cases. In the first we fix $z_\mathrm{start} =10^4$, while in the second case we let it vary (specifically, we sample over $\log_{10} z_{\mathrm{start}}$). The constraints from the different data combinations on this IC are shown in the last two blocks of Table~\ref{Tab:Vmodes_only} and Figures~\ref{fig:Vmodes_only_SMD_fix_zs} and~\ref{fig:Vmodes_only_SMD_vary_zs}. Note that we do not consider constraints from SPA alone when we sample over $\log_{10} z_{\mathrm{start}}$ since, as seen in the other cases, they are much weaker than when including $B$-modes data.

In the first case ($z_\mathrm{start} =10^4$), unlike the previous ICs, we see a slightly red-tilted spectrum, albeit one that is consistent with scale invariance. From SPA + BK18, we find $r_\mathrm{v} \ < \ 3.1 $ and $n_v\ =\ -0.36^{+0.96}_{-0.82}$. However, when we substitute BK18 with SPTpolBB, we find $r_\mathrm{v}\ = \ 4.7\pm 2.1$ (with an upper limit of 8.3) and $n_v\ =\ -0.75^{+0.55}_{-0.68}$\footnote{We tested the possibility that this behavior is due to using the \texttt{lite} version of the SPTpolBB likelihood by running the same MCMC using the multi-frequency likelihood (also implemented in \texttt{candl}). We find both chains matching, and no appreciable correlation between the \vmodes{} parameters and the nuisance ones.}. This value of \rv{} deviates from 0 at $2.2\ \sigma$, and it is the only case where we find such a deviation from $r_\mathrm{v} \ = \ 0$ among all the different cases. On the other hand, once we add BK18, we go back to just an upper limit (in this case $r_\mathrm{v} \ <\ 4.2 $).

To understand where this peak in the posterior of \rv{} originates from, let us look again at the $B$-modes plot of Figure~\ref{Fig:CMB_Spectra}. There, we can see that, for the same \rv{} and $n_v$, the SMD case results in a $BB$ spectrum of higher amplitude compared to the OCT case, specially at $\ell \gtrsim 100$. This allows the SMD case, when constrained with SPTpolBB data alone, to pass through the latter's bandpowers perfectly, as can be seen from Figure~\ref{fig:BB_spec_SMD}, whereas this is not the case for the OCT IC. Note that we can also see from Figure~\ref{fig:BB_spec_SMD} the contribution of lensing to the $B$-modes power spectrum\footnote{The lensing contribution is the same across all different cases considered in this work. This is because, at first order in perturbation theory, \vmodes{} and scalar modes do not impact each other, resulting in the same values for the remaining cosmological parameters.} which is the dominant contribution and explains why in all the considered cases \rv{} is compatible with 0. However, a large \rv{} in the SMD case would not be compatible with BK18 data, and that is why adding it to SPA + SPTpolBB results back to only an upper limit ($r_\mathrm{v}\ < \ 4.2$). That is also why we do not find any preference for this model compared to \LCDM{} with \rv{} = 0 ($\Delta\chi^2\ = \ 0$).

The second scenario of the SMD IC (where we sample over $\log_{10}(z_c)$) features consistent trends as in the previous one, with slight shifts in the upper limits and the mean values. These small shift can be explained by the fact that this IC has a strong oscillatory behavior at initial times (see Figure 7 and eq.(3.43) of~\cite{Vmodes_Paper}). This means that, depending on when the \vmodes{} were sourced, they could approach the recombination time either with a trough or a crest, which will then impact the resultant amplitude of $B$-modes.

Before moving to the case where we include \tmodes{}, it is important to stress that having \rv{} deviate from 0 at $2.2\ \sigma$ for the SMD case with SPA + SPTpolBB is not a proof of detection of \vmodes{}. First, the deviation from 0 is not statistically significant. Second, the hint goes away when including large scale data from BK18, data that is lacking in SPTpolBB. Nevertheless, this behavior specific for the SMD case merits more detailed investigation, particularly from the theory side, i.e. to find a better theoretical motivation for the presence of such sourcing of \vmodes{}. This will be a topic for future work.

\begin{table}[!htbp]
    \centering
    \renewcommand{\arraystretch}{1.2}
	\begin{tabular}{|c|c|c|c|c|}
		\hline
		Parameter & SPA & SPA + BK18 & SPA + SPTpolBB & SPA + SPTpolBB + BK18 \\
		\hline
        \multicolumn{4}{|c|}{\textbf{ISO}} & $\Delta\chi^2\ =\ -1.2$ \\
        \hline
		$r_\mathrm{v}\times 10^4$ & $< 19 $ & $< 2.7$ & $< 2.5\ \left(1.21^{+0.52}_{-0.97}\right)$ & $< 1.3$ \\ 
		\hline
		$n_v$ & $0.87^{+0.59}_{-1.0}$ & $1.7^{+1.3}_{-1.0}$ & $0.13^{+0.34}_{-0.83}$ & $0.81^{+0.69}_{-1.1}$ \\ 
		\hline
        \hline
        \multicolumn{4}{|c|}{\textbf{OCT}} & $\Delta\chi^2\ =\ -0.56$\\
        \hline
        $r_\mathrm{v}$ & $< 18$ & $< 7.2$ & $< 17$ & $< 6.8$ \\ 
		\hline
		$n_v$ & $3.5\pm 1.0$ & $4.49^{+0.87}_{-0.64}$ & $4.04^{+0.85}_{-0.73}$ & $4.47^{+0.93}_{-0.66}$ \\ 
		\hline
        \hline
        \multicolumn{4}{|c|}{\hspace{1.2cm}\textbf{SMD} $\left(z_\mathrm{start} =10^4\right)$} &  $\Delta\chi^2\ =\ 0.0$\\
        \hline
        $r_\mathrm{v}$ & $< 27$ & $< 3.1$ & $< 8.3\ \left(4.7\pm 2.1\right)$ & $< 4.2$ \\ 
		\hline
		$n_v$ & $-0.71^{+0.47}_{-0.61}$ & $-0.36^{+0.96}_{-0.82}$ & $-0.75^{+0.55}_{-0.68}$ & $-0.32^{+0.81}_{-0.67}$ \\ 
		\hline
        \hline
        \multicolumn{5}{|c|}{\textbf{SMD} (varying $z_\mathrm{start}$)}\\
        \hline
        $r_\mathrm{v}$ & $---$ & $< 2.6$ & $< 8.0\ \left(4.1^{+2.1}_{-2.7}\right)$ & $< 3.6$ \\
		\hline
		$n_v$ & $---$ & $0.01^{+1.0}_{-0.84}$ & $-0.59\pm 0.68$ & $-0.08^{+0.87}_{-0.60}$ \\
		\hline
		$\log_{10} z_{\mathrm{start}}$ & $---$ & $4.19^{+0.13}_{-0.71}$ & $4.01^{+0.19}_{-0.39}$ & $4.02^{+0.18}_{-0.48}$ \\
		\hline
	\end{tabular}
	\caption{Constraints on the amplitude of the \vmodes{} power spectrum relative to scalar perturbations (\rv) and its spectral index ($n_v$) from different data combinations (see Section~\ref{Sec:Data_Sets_Numerical_Setup}) for the three ICs considered in this work. We report the 95\% confidence level (C.L.) upper limits on \rv{} and the mean and standard deviation of $n_v$. When applicable, the 68\% confidence interval for \rv{} is given in parenthesis. The fourth case allows the SMD IC to vary the redshift at which \vmodes{} are sourced ($z_\mathrm{start}$). Except for the last case, we report the best-fit $\chi^2$ of each IC relative to $\Lambda$CDM for the SPA + SPTpolBB + BK18 combination; $\Delta\chi^2 = \chi^2_\mathrm{model} - \chi^2_{\Lambda\mathrm{CDM}}$ is indicated next to the name of each initial condition.}
    \label{Tab:Vmodes_only}
\end{table}

\subsection{Adding Tensor Modes}
\label{Sec:Res_Tmodes_Only}

We now consider the simultaneous presence of the three types of perturbations (scalar, vector and tensor) and perform similar constraints to the previous section, with some differences. First, we do not consider constraints from the SPA combination alone since, as we have seen in the previous section, it is not as constraining as when $B$-modes are included. Second, for the SMD IC, we only consider the case where $z_\mathrm{start}$ is fixed to $10^4$. Finally, we also show the constraints on $n_s$ in this section, to make a more explicit comparison between the power spectra of the three types of perturbations.

Adding \tmodes{} does not change the qualitative behavior seen in the previous case for each IC; it merely results in slight shifts in the constraints. The most up to date constraints on these perturbations from the \allData{} data combination are shown in the last column of Table~\ref{Tab:Vmodes_Tmodes}. Furthermore, the lack of statistical preference for any of the cases over the \rv{} = 0 case is evident from the values of $\Delta\chi^2$ appearing in the Table. When compared to the scenario where \vmodes{} or \tmodes{} are considered alone (see Table~\ref{Tab:Tmodes_only} for the case of \tmodes{} alone), we see here that \rv{} and $r$ decrease slightly. This is expected given that when both types of perturbations contribute to the $B$-modes power spectrum, and in order to match a given data set, the individual contributions should decrease slightly. Moreover, for the three ICs, we always have a slightly red-titled scalar power spectrum and an almost scale invariant tensor power spectrum.

For the SMD case constrained with SPA + SPTpolBB, we still see the same behavior of \rv{} as in the previous section. This can be understood by looking at Figure~\ref{Fig:CMB_Spectra}. There, we can see that the \tmodes{} contribution to the CMB $B$-modes power spectrum is $2-8$ times smaller than that of SMD in the range $500\lesssim \ell\lesssim 900$, where SPTpolBB is most constraining.

Finally, it is important to stress that the upper-limits on $r$ should not be used to draw conclusions concerning inflation models for two reasons. First, the constraints presented here implicitly assume instantaneous reheating, whereas the more realistic non-instantaneous reheating should be considered before excluding any inflation model from the upper limits quoted here~\cite{Martin:2016iqo,Zharov:2025zjg}\footnote{Implementing a non-instantaneous reheating mechanism in \texttt{CLASS} and studying the impact on the constraints presented here is beyond the scope of this work.}. Second, we are treating the origin of each IC of \vmodes{} agnostically. However, their origin could be embedded in an inflation model, which will then impact the evolution of \tmodes{}, and hence the upper limits of $r$.

\begin{table}[!htbp]
    \centering
    \renewcommand{\arraystretch}{1.2}
	\begin{tabular}{|c|c|c|c|}
		\hline
		Parameter & SPA + BK18 & SPA + SPTpolBB & SPA + SPTpolBB + BK18 \\
        \hline
        \multicolumn{3}{|c|}{\textbf{ISO}} & $\Delta\chi^2\ =\ -0.97$\\
		\hline
        $n_s$ & $0.9674\pm 0.0037$ & $0.9690\pm 0.0035$ & $0.9682\pm 0.0035$ \\ 
        \hline
		$r_\mathrm{v}\times 10^4$ & $< 2.5$ & $< 2.4$ & $< 1.2$ \\ 
		\hline
		$n_v$ & $2.0^{+1.2}_{-0.97}$ & $0.28^{+0.35}_{-0.95}$ & $1.2^{+0.77}_{-1.2}$ \\ 
		\hline
		$r$ & $< 0.032$ & $< 0.13$ & $< 0.031$ \\ 
		\hline
		$n_\mathrm{t,scc}\times 10^3$ & $-1.8^{+1.6}_{-0.61}$ & $-6.1^{+6.1}_{-1.6}$ & $-1.7^{+1.6}_{-0.53}$ \\ 
        \hline
        \hline
        \multicolumn{3}{|c|}{\textbf{OCT}} & $\Delta\chi^2\ =\ -0.93$\\
		\hline
        $n_s$ & $0.9686\pm 0.0035$ & $0.9686\pm 0.0035$ & $0.9682\pm 0.0035$ \\ 
        \hline
		$r_\mathrm{v}$ & $< 6.3$ & $< 21$ & $< 6.3$ \\ 
		\hline
		$n_v$ & $4.6^{+0.90}_{-0.73}$ & $4.1^{+0.84}_{-0.66}$ & $4.7^{+0.84}_{-0.62}$ \\ 
		\hline
		$r$ & $< 0.031$ & $< 0.12$ & $< 0.031$ \\ 
		\hline
		$n_\mathrm{t,scc}\times 10^3$ & $-1.8^{+1.6}_{-0.58}$ & $-6.1^{+6.0}_{-1.8}$ & $-1.7^{+1.6}_{-0.56}$ \\ 
        \hline
        \hline
        \multicolumn{3}{|c|}{\textbf{SMD} $\left(z_\mathrm{start}=10^4\right)$} & $\Delta\chi^2\ =\ -0.39$\\
		\hline
        $n_s$ & $0.9690\pm 0.0035$ & $0.9698\pm 0.0035$ & $0.9690\pm 0.0034$ \\ 
        \hline
		$r_\mathrm{v}$ & $< 3.2$ & $< 8.5\ \left(4.8\pm 2.2\right)$ & $< 4.0$ \\ 
		\hline
		$n_v$ & $-0.16^{+0.92}_{-0.75}$ & $-0.72^{+0.53}_{-0.63}$ & $-0.18^{+0.76}_{-0.59}$ \\ 
		\hline
		$r$ & $< 0.032$ & $< 0.12$ & $< 0.032$ \\ 
		\hline
		$n_\mathrm{t,scc}\times 10^3$ & $-1.9^{+1.6}_{-0.63}$ & $-6.1^{+5.9}_{-1.8}$ & $-1.8^{+1.6}_{-0.63}$ \\ 
		\hline
\end{tabular}
	\caption{Same as Table~\ref{Tab:Vmodes_only} but including \tmodes{}.}
    \label{Tab:Vmodes_Tmodes}
\end{table}

\begin{table}[!htbp]
\begin{tabular}{|c|c|c|c|}
\hline
Parameter & SPA + BK18 & SPA + SPTpolBB & SPA + SPTpolBB + BK18 \\
\hline
\multicolumn{3}{|c|}{\textbf{\tmodes{}}} & $\Delta\chi^2\ =\ -0.12$\\
\hline
$n_s$ & $0.9684 \pm 0.0035$ & $0.9686 \pm 0.0035$ & $0.9678 \pm 0.0035$ \\
\hline
$r$ & $< 0.034 $ & $< 0.13$ & $< 0.034 $ \\ 
\hline
$n_\mathrm{t,scc}\times 10^3$ & $-2.0^{+1.6}_{-0.76}$ & $-6.4^{+6.3}_{-1.9}$ & $-2.0^{+1.6}_{-0.74}$ \\ 
\hline
\end{tabular}
\caption{Same as Table~\ref{Tab:Vmodes_Tmodes} but for \tmodes{} only.}
\label{Tab:Tmodes_only}
\end{table}

\section{Conclusion}
\label{Sec:Conclusion}
In this work, we present novel constraints on gravitational Vector modes (\vmodes{}) sourced with three different initial conditions (ICs). The first, called neutrino isocurvature (ISO), stems from a velocity difference between photons and neutrinos in the primordial universe. This velocity difference creates an anisotropic stress that sources \vmodes{}. The second IC, dubbed neutrino octupole (OCT), sources \vmodes{} by assuming a non-vanishing primordial value of the neutrino octupole. The latter will source the neutrino quadrupole, which then sources the \vmodes{}. The last IC, named sourced mode (SMD), assumes the presence of an anisotropic stress that sharply sources \vmodes{} at a specific redshift ($z_\mathrm{start}$) before matter-radiation equality. Our baseline SMD model has $z_\mathrm{start} = 10^4$, but we also consider the situation where this is a free parameter. Finally, we also consider the presence of tensor perturbations (\tmodes{}), either alone or with \vmodes{} for each of these ICs.

Using CMB $TT$, $TE$ and $EE$ data from \sptlr{}, \ACTDR{} and \planck{} (collectively abbreviated as SPA), along with $BB$ data from BICEP/Keck (BK18) and SPTpol, we find no statistically significant evidence for \vmodes{}. However, their presence is not definitely excluded, and the most up to date upper limits on their amplitude with SPA + SPTpolBB + BK18 are $r_\mathrm{v} < 1.3\times10^{-4}$, $r_\mathrm{v} < 6.8$, and $r_\mathrm{v} < 4.2$ for the three ICs, respectively. We also find that ISO and SMD ICs have, respectively, a slightly blue-tilted and red-tilted spectra, though they are consistent with scale invariance. On the other hand, the OCT IC shows a highly blue-tilted spectrum, deviating from scale invariance at $\sim\ 7\sigma$. When adding \tmodes{}, we find a slightly lower upper limits on \rv{} and $r$ (the amplitude of \tmodes{}) for each IC, and no appreciable change in the spectral index of \vmodes{}, $n_v$.

When constraining SMD with SPA + SPTpolBB, we find $r_\mathrm{v}\ =\ 4.7\pm 2.1$ ($4.8\pm 2.2$ when \tmodes{} are included), which is the only case considered here that shows such a constraint. However, this is consistent with $r_\mathrm{v}=0$ at the $\sim2\sigma$ level, i.e. consistent with being a statistical fluctuation above 0, and when BK18 is included such a constraint goes back to being just an upper limit $r_\mathrm{v}<4.2$ ($<4.0$ when \tmodes{} are included).

Our results have two important implications. First, they reinforce the importance of CMB $B$-mode measurements in constraining the fundamental properties of the Universe, further motivating upcoming measurements by LiteBIRD~\cite{LiteBIRD:2020khw,LiteBIRD:2022cnt}, BICEP Array~\cite{Bicep_Future_1,BICEP_Future_2}, and the South Pole Telescope. Second, they indicate that \vmodes{} remain an interesting subject of investigation, particularly in the context of fundamental theories, such as modified gravity, loop quantum gravity or brane-world cosmology~\cite{Brandenberger_Vmodes,Bouncing1,Bouncing2,LQG1,Brane_World_Roy,Vector_Inf1,Vector_Inf2}. This, in turn, opens the possibility of testing these theories through the imprint of their \vmodes{} on the CMB.


\subsection*{Acknowledgments}

Special thanks and appreciation to Shohei Saga, Kaito Yura, Shuichiro Yokoyama and Kiyotomo Ichiki who contacted us during the final stages of this work, informing us of a similar analysis they were conducting. This allowed both teams to compare and confirm their analysis settings, which helped in concretizing the conclusions reached by both teams. We also thank Silvia Galli and Lennart Balkenhol from IAP for fruitful discussions during all stages of this project. This project has received funding from the European Research Council (ERC) under the European Union’s Horizon 2020
research and innovation program (grant agreement No 101001897). This work has also made use of the Infinity Cluster hosted by IAP. We thank Stephane Rouberol for smoothly running this cluster for us.


\bibliography{Biblio.bib}

\end{document}